\documentclass[10pt, final, twocolumn, journal]{IEEEtran}

\usepackage{makeidx}
\makeindex

 \usepackage[pdftex]{graphicx}
 \usepackage{epstopdf}
\usepackage[usenames]{color}

\usepackage[color,final]{showkeys}

\usepackage{amsmath,amssymb,amsthm}
\usepackage{latexsym,amsfonts,amscd,amsxtra,amstext}
\usepackage{algorithm,algorithmic}
\usepackage{url}

\usepackage{epsfig} % for Zhu Han's images
\usepackage{threeparttable}
\usepackage{index}
\usepackage{subfigure}

\usepackage[normalem]{ulem} % to use \sout

\newcommand{\cB}{{\mathcal{B}}}

\newcommand{\RR}{\mathbb{R}}

\newcommand{\sign}{\mathrm{sign}}
\newcommand{\vzero}{\mathbf{0}}
\newcommand{\vone}{{\mathbf{1}}}

\newcommand{\supp}{{\mathrm{supp}}} % support
\newcommand{\bc}{\begin{center}}
\newcommand{\ec}{\end{center}}

\newcommand{\bdm}{\begin{displaymath}}
\newcommand{\edm}{\end{displaymath}}

\newcommand{\beq}{\begin{equation}}
\newcommand{\eeq}{\end{equation}}

\newcommand{\bfl}{\begin{flushleft}}
\newcommand{\efl}{\end{flushleft}}

\newcommand{\bt}{\begin{tabbing}}
\newcommand{\et}{\end{tabbing}}

\newcommand{\beqn}{\begin{eqnarray}}
\newcommand{\eeqn}{\end{eqnarray}}

\newcommand{\beqs}{\begin{align*}} % no equation numbers
\newcommand{\eeqs}{\end{align*}}  % no equation numbers

\newtheorem{theorem}{Theorem}
\newtheorem{condition}{Condition}
\newtheorem{assumption}{Assumption}

\newtheorem{corollary}{Corollary}
\newtheorem{remark}{Remark}
\newtheorem{lemma}{Lemma}
\ifCLASSINFOpdf
\else
\fi

\begin{document}

\title{Necessary and sufficient conditions of solution uniqueness in $\ell_1$ minimization}

\author{
\thanks{The work of H. Zhang is supported by the China Scholarship Council during his visit to Rice University, and in part by the Graduate School of NUDT under Funding of Innovation B110202 and  by Hunan Provincial Innovation Foundation For Postgraduate CX2011B008. The work of W. Yin is supported in part by  NSF grants DMS-0748839 and ECCS-1028790, and ONR Grant N00014-08-1-1101. The work of L. Cheng is supported by the National Science Foundation of China under Grants No. 61271014 and No.61072118. }
{Hui~Zhang,~
        Wotao~Yin,
        and~Lizhi~Cheng}% <-this % stops a space
\thanks{Hui Zhang and Lizhi Cheng are with the Department of Mathematics and Systems Science,
College of Science, National University of Defense Technology,
Changsha, Hunan,  China. 410073.  Emails: \texttt{hui.zhang@rice.edu} and \texttt{clzcheng@nudt.edu.cn} }% <-this % stops a space
\thanks{Wotao Yin is with the Department of Computational and Applied Mathematics, Rice University, Houston, Texas, USA. 77005. Email: \texttt{wotao.yin@rice.edu} }% <-this % stops a space
}

\markboth{Journal of \LaTeX\ Class Files,~Vol.~?, No.~?, January~????}%
{Shell \MakeLowercase{\textit{et al.}}: Bare Demo of IEEEtran.cls for Journals}

\maketitle

\begin{abstract}
%\boldmath
This paper shows that the solutions to various convex $\ell_1$ minimization problems are \emph{unique} if and only if a common set of  conditions are satisfied.  This result applies broadly to the basis pursuit model, basis pursuit denoising model, Lasso model, as well as other $\ell_1$ models that either minimize $f(Ax-b)$ or impose the constraint $f(Ax-b)\leq\sigma$, where $f$ is a strictly convex function. For these models, this paper proves that,  given a solution $x^*$ and defining  $I=\supp(x^*)$ and $s=\sign(x^*_I)$, $x^*$ is the unique solution if and only if $A_I$ has full column rank and there exists $y$ such that $A_I^Ty=s$ and $|a_i^Ty|_\infty<1$ for $i\not\in I$. This condition is previously known to be sufficient for the basis pursuit model to have a unique solution supported on $I$. Indeed, it is also necessary, and applies to a variety of other $\ell_1$ models. The paper also discusses ways to recognize unique solutions and verify the uniqueness conditions numerically.
\end{abstract}

\begin{IEEEkeywords}
 basis pursuit, $\ell_1$ minimization, solution uniqueness, strict complementarity
\end{IEEEkeywords}

\IEEEpeerreviewmaketitle

\section{Introduction}
Let $x\in\RR^n$ be the decision variable. This paper studies the unique solutions of the  $\ell_1$ minimization problems  including the basis pursuit problem  \cite{CDS}
\begin{subequations}
\label{eq:qp}
\beq\label{eq:bp}
\min\|x\|_1\quad\text{s.t.}~Ax=b
\eeq
and  convex problems
\begin{align}
\label{eq:qpa}
&\min f_1(Ax-b)+\lambda \|x\|_1,\\
\label{eq:qpb}
&\min \|x\|_1,\quad\text{s.t.}~f_2(Ax-b)\leq \sigma, \\
\label{eq:qpc}
&\min f_3(Ax-b), \quad\text{s.t.}~  \|x\|_1\leq \tau,
\end{align}
\end{subequations}
where $\lambda,\sigma,\tau>0$ are scalar parameters,  $A$ is a matrix, and $f_i(x), i=1, 2, 3$ are strictly convex functions. The Lasso problem \cite{Ti} is a special case of  problem \eqref{eq:qpa} or \eqref{eq:qpc} while the basis pursuit denoising problem \cite{CDS} is a special case of  problem \eqref{eq:qpb} all with $f_i(\cdot)=\frac{1}{2}\|\cdot\|_2^2,~ i=1, 2, 3$.

There is a rich literature on analyzing, solving, and applying problems in forms of \eqref{eq:bp}--\eqref{eq:qpc} in the communities of information theory, signal processing, statistics,  machine learning, optimization, and so on. In many cases, problems \eqref{eq:bp}--\eqref{eq:qpc} need to have unique solutions; when there are more than one solution, the set of solutions is a convex set including an infinite number of solutions. In compressive sensing  signal recovery, having non-unique solutions means that the underlying signal can not be reliably recovered from the given data.  In feature selection, non-unique solutions cause ambiguity for feature identification. Even certain optimization methods and algorithms, especially those producing the solution path of \eqref{eq:qpa}--\eqref{eq:qpc} over varying parameters such as LARS \cite{EJHT} and parametric quadratic programming \cite{BG}, require solution uniqueness; they will fail (or need special treatments) upon encountering non-unique solutions. Therefore, establishing a condition of solution uniqueness is important for both the analysis and  computation of problems \eqref{eq:bp}--\eqref{eq:qpc}.  Fortunately, there are various sufficient conditions guaranteeing solution uniqueness in problem \eqref{eq:bp}, such as  Spark \cite{DE, BDE},  the mutual incoherence condition \cite{DH,EB}, the null-space property (NSP) \cite{Do06, Co}, the restricted isometry principle (RIP) \cite{CT}, the spherical section property \cite{Zh}, the ``RIPless" property \cite{CP11}, and so on. Some conditions guarantee the unique recovery of a given solution or solutions with a given set of signs; other conditions provide the guarantees for all solutions with sufficiently few nonzero entries. However, none of them is known to be both necessary and sufficient for solution uniqueness in problems  \eqref{eq:qpa}--\eqref{eq:qpc}.  This paper shows that  given a solution $x^*$ to any problem among \eqref{eq:bp}--\eqref{eq:qpc}, Condition \ref{cond:unique} below is  both necessary and sufficient for $x^*$ to be the unique solution. Hence, it is weaker than the  sufficient conditions listed above.

We let $X$, $X_\lambda$, $Y_\sigma$, and $Z_\tau$ denote the sets of solutions to  problems \eqref{eq:bp}--\eqref{eq:qpc}, respectively. We let $a_i$ be the $i$th column of $A$ and $x_i$ be the $i$th entry of $x$. Given an index set $I$, we frequently use $A_I$ as the submatrix of $A$ formed by its columns $a_i$,  $i\in I$ and  $x_I$ as the subvector of $x$ formed by entries $x_i$, $i\in I$.

Our analysis makes the following assumptions:

\begin{assumption}\label{assmp1}
Matrix $A$ has full row rank.
\end{assumption}
\begin{assumption}\label{assmp2}
The solution sets $X$, $X_\lambda$, $Y_\sigma$, and $Z_\tau$ of problems \eqref{eq:bp}--\eqref{eq:qpc}, respectively, are nonempty.
\end{assumption}
\begin{assumption}\label{assmp3} In problems \eqref{eq:qpa}--\eqref{eq:qpc}, functions $f_1,f_2,f_3$ are strictly convex. In addition,  the constraint of problem \eqref{eq:qpc} is bounding, namely, $\tau \le \inf \{\|x\|_1: f_3(Ax-b) = f_3^*\}$, where $f_3^*:=\min_{y\in\RR^n} f(Ay- b)$.
\end{assumption}
Assumptions \ref{assmp1} and \ref{assmp2} are standard. If Assumption \ref{assmp1} does not hold and $Ax=b$ is consistent, the problems can be simplified; specifically, one can decompose $A=\begin{bmatrix}A_1\\ A_2\end{bmatrix}$ and $b=\begin{bmatrix}b_1\\ b_2\end{bmatrix}$ so that and  $A_1$ has full row rank equal to $\mathrm{rank}(A)$, and one can replace the constraints $Ax = b$ by $A_1x = b_1$ and introduce functions $\bar{f}_i$ so that $\bar{f}_i(A_1x - b_1) \equiv f_i(Ax-b)$, $i=1,2,3$. Assumption \ref{assmp2} guarantees that the solutions of problems \eqref{eq:bp}--\eqref{eq:qpc} can be attained so the discussion of solution uniqueness makes sense.  The strict convexity of $f_1,f_2,f_3$ and the restriction on $\tau$ in Assumption \ref{assmp3} are  also fairly basic toward solution uniqueness. Strict convexity rules out piece-wise linearity. (Note that $f_1,f_2,f_3$ can still be non-differentiable.) If the restriction on $\tau$ is removed, the solution uniqueness of  problem \eqref{eq:qpc} becomes solely  up to $f_3(Ax-b)$, independent of $\|x\|_1$. % and Figure \ref{?} illustrates this point by an example.

For a given vector $x^*$, the following conditions on matrix $A$ is the key to solution uniqueness, and its sufficiency has been established in \cite{Fu}.

\begin{condition}\label{cond:unique}  Under the definitions  $I:=\supp(x^*)\subseteq\{1,\ldots,n\}$ and $s:=\sign(x_I^*)$, matrix $A\in\mathbb{R}^{m\times n}$ has the following properties:
\begin{enumerate}
\item submatrix $A_{I}$ has full column rank, and
\item there is  $y \in \mathbb{R}^m$ obeying  $A_{I}^Ty=s$ and $\|A_{I^c}^Ty\|_\infty <1$.
\end{enumerate}
\end{condition}
The main theorem of this paper asserts that Condition \ref{cond:unique} is both necessary and sufficient to the uniqueness of solution $x^*$.
\begin{theorem}[Solution uniqueness]\label{thm:unique} Under Assumptions \ref{assmp1}--\ref{assmp3}, given that $x^*$ is a solution  to problem \eqref{eq:bp}, \eqref{eq:qpa}, \eqref{eq:qpb}, or \eqref{eq:qpc}, $x^*$ is  the unique solution if and only if Condition \ref{cond:unique} holds.
\end{theorem}
In addition, combining Theorem \ref{thm:unique} with the optimality conditions for problems \eqref{eq:bp}--\eqref{eq:qpc}, the following theorems give the necessary and sufficient conditions of \emph{unique optimality} for those problems.
\begin{theorem}[Basis pursuit  unique optimality]\label{thm:bp} Under Assumptions \ref{assmp1}--\ref{assmp2},
$x^*\in\mathbb{R}^n$ is the unique solution to problem \eqref{eq:bp} if and only if $Ax^*=b$ and Condition \ref{cond:unique} is satisfied.
\end{theorem}
\begin{theorem}[Problems \eqref{eq:qpa}--\eqref{eq:qpc} unique optimality]\label{thm:qp}
Under Assumptions \ref{assmp1}--\ref{assmp3} and the additional assumption $f_1,f_2,f_3\in C^1$, $x^*\in\mathbb{R}^n$ is the unique solution to problem \eqref{eq:qpa}, \eqref{eq:qpb}, or \eqref{eq:qpc} if and only if, respectively,
\begin{subequations}\label{eq:opt}
\begin{align}
\label{eq:opta}
&\exists~ p^*\in \partial\|x^*\|_1, \ni
p^*+\lambda A^T \nabla f_1(Ax^*-b) = \vzero,\\ \nonumber \\
\label{eq:optb}\nonumber
&f(Ax^*-b)\le \sigma\quad\text{and}\quad \exists~ p^*\in \partial\|x^*\|_1, \eta\ge 0,\\
&\ni p^*+\eta A^T \nabla f_2(Ax^*-b) = \vzero,~\text{or}\\
 \nonumber \\
\label{eq:optc}\nonumber
&\|x^*\|_1\le \tau\quad\text{and}\quad  \exists~p^*\in \partial\|x^*\|_1, \nu\ge 0,\\
& \ni \nu p^*+ A^T \nabla f_3(Ax^*-b) = \vzero,
\end{align}
\end{subequations}

and in addition Condition \ref{cond:unique} holds.
\end{theorem}
The proofs of these theorems are given in Section \ref{sc:proofs} below.

\subsection{Related works}  Since the sufficiency is not the focus of this paper, we do not go into more details of the  sufficient conditions that have been mentioned above. We just point out that several papers such as \cite{Fu,CR11} construct the least-squares (i.e., minimal $\ell_2$-norm) solution $\bar{y}$ of $A_{I}^Ty=s$ and establish  sufficient conditions for $\|A_{I^c}^T\bar{y}\|_\infty <1$ to hold. Next, we review the existing  results on  necessary conditions for the  uniqueness of $\ell_1$ minimizer.

Work \cite{CRT} considers  problem \eqref{eq:bp} with complex-valued quantities and  $A$ equal to a down-sampled discrete Fourier operator, for which it establishes both the necessity and  sufficiency of Condition 1 to the solution uniqueness of  \eqref{eq:bp}. Their proof uses the Hahn-Banach separation theorem and the Parseval formula. Work \cite{Tr} lets the entries of matrix $A$ and vector $x$ in problem \eqref{eq:bp}  have complex values and gives a  sufficient condition for its solution uniqueness.  In regularization theory, Condition 1 is used to derive linear error bounds under the name of range or source conditions  in \cite{GHS}, which shows the necessity and sufficiency of Condition 1 for solution uniqueness of \eqref{eq:bp} in a Hilbert-space setting.  More recently,   \cite{Do} constructs the set
\begin{eqnarray}\nonumber
\mathbb{F}=\{x:  \|A_{J^c}^TA_{J}(A^T_{J}A_{J})^{-1}\sign(x_{J})\|_\infty<1 ~\textrm{and}~\\
\textrm{rank}(A_J)=|J|\},~\text{where}~J=\textrm{supp}(x), \nonumber
\end{eqnarray}
and then states that the set of vectors that can be recovered by problem \eqref{eq:bp} is exactly characterized by the closure of  $\mathbb{F}$ if the measurement matrix $A$ satisfies the so-called \emph{general position} (GP) condition, namely, for any sign vector $s\in\{-1,1\}^n$, the set of columns $\{A^i\}$ of $A\in\RR^{m\times n}$ satisfying that any $k$-dimensional affine subspace of $\RR^m$, $k< m$, contains at most $k+1$ points from the set $\{s_iA^i\}$.
This paper claims that the result holds without the GP condition.

To our knowledge, there are few conditions addressing the solution uniqueness of problems \eqref{eq:qpa}--\eqref{eq:qpc}. The following conditions  in \cite{Fu}, \cite{Fu05} are sufficient for $x^*$ to the unique minimizer of \eqref{eq:qpa} for $f_1(\cdot)=\frac{1}{2}\|\cdot\|_2^2$:
\begin{subequations}\label{cond0}
\begin{align}
&A_{I}^T(b-A_Ix^*_I)=\lambda\cdot \sign(x_I^*),\label{cond2a}\\
 &\|A_{I^c}^T(b-A_Ix^*_I)\|_\infty <\lambda, \label{cond2b}\\
 & A_I~\textrm{has~full~column~rank.}
\end{align}
\end{subequations}
%which was derived by distinguishing the nonzero components and zero components of $x^*$ in the typical necessary and sufficient condition (2a) and taking a strict inequality to get (\ref{cond2b}).
However, they are not necessary as demonstrated by the following example.
Let 
\beq\label{exam}
A=\begin{bmatrix}1 & 0& 2\\ 0 & 2 &-2 \end{bmatrix},~b=\begin{bmatrix}1\\ 1\end{bmatrix}, ~\lambda =1
\eeq
and consider solving the Lasso problem, which is a special case of problem \eqref{eq:qpa}:
\beq\label{lasso}
\min \frac{1}{2}\|Ax-b\|_2^2+\lambda \|x\|_1.
\eeq
One gets the \emph{unique} solution $x^* = [0~1/4~ 0]^T$ and $I=\supp(x^*)=\{2\}$. However, the inequality in condition (\ref{cond2b}) holds with equality. In general, conditions \eqref{cond0} becomes necessary in case $A_I$ happens to be a full rank square matrix. This assumption, however, does not apply to a sparse solution $x^*$.
 Nevertheless, we summarize the result in the following corollary, whose proof is given at the end of Section \ref{sc:proofs}.
\begin{corollary}\label{corol}
If $x^*$ is the unique minimizer of problem \eqref{eq:qpa} with $f_1(\cdot)=\frac{1}{2}\|\cdot\|_2^2$ and if $A_{I}$, where $I=\supp(x^*)$, is a square matrix with full rank, then condition (\ref{cond0}) holds.
\end{corollary}

Very recently, work \cite{Ti12} investigates the solution uniqueness of 
\eqref{lasso} and presents the following result.
\begin{theorem}[\cite{Ti12}]\label{Ti12}
 Let $x^*$ be a solution of \eqref{lasso} and $J:=\{i: |a_i^T(b-Ax^*)|=\lambda\}$. If submatrix $A_J$ is full column rank, then $x^*$ is unique. Conversely, for \emph{almost} every $b\in \RR^m$, if  $x^*$ is unique, then $A_J$ is full column rank.
\end{theorem}
In Theorem \ref{Ti12},  the necessity part ``for almost every $b$'' is new. Indeed, it is not for every $b$. An example is given in (\ref{exam}) with a unique solution $x^*$ and  $J=\{1, 2, 3\}$, but   $A_J$ does not full column rank. On the other hand, we can figure out a special case in which the full column rankness of $A_J$ becomes necessary for all $b$ in the following corollary, whose proof is  given at the end of Section \ref{sc:proofs}.

\begin{corollary}\label{coro2}
Let $x^*$ be a solution of problem \eqref{lasso} and define $I:=\supp(x^*)$ and $J:=\{i: |a_i^T(b-Ax^*)|=\lambda\}$. If $|J|=|I|+1$, then  $x^*$ is the unique solution if and only if $A_J$ has full column rank.
\end{corollary}

\section{Proofs of Theorems \ref{thm:unique}--\ref{thm:qp}}\label{sc:proofs}
We establish Theorem \ref{thm:unique} in three steps. Our first step proves the theorem for problem \eqref{eq:bp} only. Since the only difference between this part and Theorem \ref{thm:bp} is the conditions $Ax^*=b$, we  prove Theorem \ref{thm:bp} first. In the second step, for problems \eqref{eq:qpa}--\eqref{eq:qpc}, we show that both $\|x\|_1$ and $Ax-b$ are constant for $x$ over the solution sets $X_\lambda, Y_\sigma,Z_\tau$, respectively. Finally,  we prove Theorem \ref{thm:unique}  for problems \eqref{eq:qpa}--\eqref{eq:qpc}.
\begin{IEEEproof}[Proof of Theorem \ref{thm:bp}] We frequently use the notions  $I=\supp(x^*)$ and $s=\sign(x_I^*)$ below.
\medskip

``$\Longleftarrow$''. This part has been shown in \cite{Fu, Tr}. For completeness, we give a  proof. Let $y$ satisfy Condition \ref{cond:unique}, part 2, and let $x\in\RR^n$ be an arbitrary vector satisfying $Ax = b$ and $x\not=x^*$. We shall show $\|x^*\|_1<\|x\|_1$.

Since $A_I$ has full column rank and $x\not=x^*$, we have $\supp(x)\not=I$; otherwise from $A_I x^*_I =b= A_I x_I$, we would get $x_I^* = x_I$ and  thus the contradiction $x^*=x$.

From $\supp(x)\not=I$, we  get $b^T y < \|x\|_1$. To see this, let $J := \supp(x) \setminus I$, which is a non-empty subset of $I^c$. From Condition \ref{cond:unique}, we have $\|A_I^T y\|_\infty =1$ and $\|A_J^T y\|_\infty < 1$, and thus
\begin{align*}
\langle x_I, A_I^T y\rangle&\le \|x_I\|_1\cdot\|A_I^T y\|_\infty\le\|x_I\|_1,\\
\langle x_J, A_J^T y\rangle&\le \|x_J\|_1\cdot\|A_J^T y\|_\infty<\|x_J\|_1,
\end{align*}
(the last inequality is ``$<$'' not ``$\le$'') which lead to
\begin{eqnarray*}
b^T y = \langle x, A^T y\rangle& = &\langle x_I, A_I^T y\rangle+ \langle x_J, A_J^T y\rangle\\
 &<& \|x_I\|_1  +\|x_J\|_1 = \|x\|_1.
\end{eqnarray*}

On the other hand, we have
$$
\|x^*\|_1 = \langle x^*_I, \sign(x^*_I)\rangle = \langle x^*_I, A_I^T y\rangle = \langle A_I x^*_I,y\rangle = b^T y
$$
and thus $\|x^*\|_1=b^T y<\|x\|_1$.%, from which it follows $x^*$ is the unique solution to \eqref{eq:bp}.
\medskip

\noindent``$\Longrightarrow$''. Assume that $x^*$ is the unique solution to \eqref{eq:bp}. Obviously, $Ax^*=b$.

It is easy to obtain Condition \ref{cond:unique}, part 1. Suppose it does not hold. Then, $A_I$ has a nontrivial null space, and perturbing $x_I^*$ along the null space will change the objective $\|x_I^*\|_1 = s^T x^*_I$ while maintaining $A_I x_I^*=b$; hence, this  perturbing breaks the unique optimality of $x^*$.
(In more details, there exists a \emph{nonzero} vector $d\in\RR^{n}$ such that $A_I d_I = \vzero$ and $d_{I^c}=\vzero$. For any scalar $\alpha$ near zero, we have $\sign(x^*_I+\alpha d_I)=\sign(x_I^*)=s$ and thus $\|x^*+\alpha d\|_1= s^T (x^*_I+\alpha d_I)=s^T x^*_I+\alpha( s^T d_I)=\|x^*\|_1+\alpha( s^T d_I)$. Since $x^*$ is the unique solution, we must have $\|x^*_I+\alpha d\|_1> \|x^*\|_1$ or, equivalently, $\alpha( s^T d_I)>0$ whenever $\alpha\not=0$. This is impossible as we can perturb $\alpha$ around 0 both ways.)

It remains to construct a vector $y$ for Condition \ref{cond:unique}, part 2.
Our construction is based on  the strong convexity relation between a linear program (called the primal problem) and its dual problem, namely, if one problem has a solution, so does the other, and the two solutions must give the same objective value. (For the interested reader, this result follows from the Hahn-Banach separation theorem, also from the theorem of alternatives \cite{Da}. Alternatively, it can be  obtain constructively via the Simplex method; specifically, whenever a primal solution exists, the Simplex method terminates in a finite number of steps with a primal-dual solution pair.)

The strong duality relation holds between  \eqref{eq:bp} and its dual problem
\beq\label{eq:bpdual}
\max_{p\in\RR^m} b^Tp\quad\text{s.t.}~ \|A^Tp\|_\infty \le 1
\eeq
because  \eqref{eq:bp} and \eqref{eq:bpdual}, as a primal-dual pair, are equivalent to the primal-dual linear programs
\begin{subequations}\label{eq:pdpair}
\begin{align}
\label{eq:pdpaira}
\min_{u,v\in\RR^n} \vone^Tu + \vone^T v & \quad \text{s.t.}~A u - Av = b, ~u\ge 0, ~v\ge 0,\\
\label{eq:pdpairb}
\max_{q\in \RR^m} b^T q& \quad \text{s.t.} -\vone\le A^T q\le \vone,
\end{align}
\end{subequations}
respectively, where the strong duality relation holds between \eqref{eq:pdpaira} and \eqref{eq:pdpairb}. By ``equivalent'', we mean that one can obtain solutions from each other by the rules:
$$\begin{array}{ll}
\text{given}~u^*,v^*, & \text{obtain}~ x^* = u^* - v^*\\
\text{given}~x^*, & \text{obtain}~ u^*= \max(x^*,\vzero), ~v^*= \max(-x^*,\vzero),\\
\text{given}~q^*, & \text{obtain}~ p^*= q^*,\\
\text{given}~p^*, & \text{obtain}~ q^*= p^*.
\end{array}$$
Therefore, since \eqref{eq:bp} has solution $x^*$, there exists a solution $y^*$ to \eqref{eq:bpdual}, which satisfies $\|x^*\|_1 = b^T y^*$ and $\|A^T y^*\|_\infty\le 1$. (One can obtain such $y^*$ from the  Hahn-Banach separation theorem or the theorem of alternatives rather directly.) However,  $y^*$ may \emph{not} obey $\|A_{I^c}^Ty^*\|_\infty <1$. We shall perturb $y^*$ so that $\|A_{I^c}^Ty^*\|_\infty <1$.

To prepare for the perturbation, we let $L:=\{i\in I^c:a_i^Ty^* = -1 \}$ and $U:=\{i\in I^c:a_i^Ty^* = 1\}$. Our goal is to perturb $y^*$ so that  $-1<a_i^Ty^*<1$ for $i\in L\cup U$ and $y^*$ remains optimal to  \eqref{eq:bpdual}.
To this end, consider for a fixed $\alpha>0$ and $t:=\|x^*\|_1$, the linear program
\beq\label{eq:iprimal}
\min_{x\in\RR^n}  \sum_{i\in L}\alpha x_i-\sum_{i\in U}\alpha x_i,\quad\text{s.t.}~Ax=b,~\|x\|_1\le t.
\eeq
Since $x^*$ is the unique solution to \eqref{eq:bp}, it is  the unique feasible solution to problem \eqref{eq:iprimal}, so problem \eqref{eq:iprimal} has the optimal objective value $\sum_{i\in U}\alpha x_i^* - \sum_{i\in L}\alpha x_i^*=0$.
By setting up equivalent linear programs like what has been done for \eqref{eq:bp} and \eqref{eq:bpdual} above,  the strong duality relation holds between \eqref{eq:iprimal} and its dual problem
\beq\label{eq:idual}
\max_{p\in\RR^m,q\in \RR} b^T p - tq,\quad\text{s.t.}~\|A^Tp - \alpha r\|_\infty \le q,~q\ge 0,
\eeq
where  $r\in\RR^n$ is given by
$$r_i = \begin{cases}1,&i\in L,\\
-1,&i\in U,\\
0,&\text{otherwise}.
\end{cases}$$
Therefore, \eqref{eq:idual} has a solution $(p^*,q^*)$ satisfying $b^T p^*- tq^*= 0$.

According to the last constraint of \eqref{eq:idual}, we have $q^*\ge 0$, which we split into two cases:  $q^*=0$ and  $q^*>0$.
\begin{itemize}
\item[i)] If $q^*=0$, we have $A^T p^*=\alpha r$ and $b^T p^*=0$.
\item[ii)] If $q^*>0$, we let $r^*:=p^*/q^*$, which satisfies $b^T r^* =t=  \|x^*\|_1$ and $\|A^Tr^* - \frac{\alpha}{q^*}r\|_\infty \le 1$, or equivalently, $-\vone+\frac{\alpha}{q^*}r\le A^Tr^*  \le \vone+\frac{\alpha}{q^*}r$.
\end{itemize}
Now we perturb $y^*$. Solve \eqref{eq:idual}  with a sufficiently small $\alpha > 0$ and obtain a solution $(p^*,q^*)$. If case i) occurs, we let $y^* \gets y^* +p^*$; otherwise, we let $y^* \gets \frac{1}{2}(y^* +r^*)$. In both cases,
\begin{itemize}
\item $b^T y^*$ is unchanged, still equal to $\|x^*\|_1$;
\item $-1<a_i^Ty^*<1$ holds for $i\in L\cup U$ after the perturbation;
\item for each $i \not \in L\cup U$, if $a_j^Ty^*\in[-1,1]$ or $a_j^Ty^*\in(-1,1)$ holds before the perturbation, the same holds after the perturbation;
\end{itemize}
Therefore, after the perturbation, $y^*$ satisfies:
\begin{itemize}
\item[1)] $b^T y^* = \|x^*\|_1$,
\item[2)] $\|A_I^T y^*\|_\infty\le 1$, and
\item[3)] $\|A_{I^c}^T y^*\|_\infty <1$.
\end{itemize}
From 1) and 2) it follows
\begin{itemize}
\item[4)] $A_I^T y = \sign(x_I^*)$

\end{itemize}
since $\|x_I^*\|_1 = \|x^*\|_1 = b^T y^* = \langle A_I x_I^*,y^*\rangle = \langle x_I^*, A_I^T y^*\rangle\le \|x_I^*\|_1 \|A^T y^*\|_\infty \le \|x_I^*\|_1$ and thus $\langle x_I^*, A_I^T y^*\rangle = \|x_I^*\|_1$, which dictates 4). From 3) and 4), Condition \ref{cond:unique}, part 2, holds with $y=y^*$.
\end{IEEEproof}

\begin{IEEEproof}[Proof of Theorem \ref{thm:unique} for problem \eqref{eq:bp}]
The above proof for Theorem \ref{thm:bp} also serves the proof of  Theorem \ref{thm:unique} for problem \eqref{eq:bp} since $Ax^*=b$ is  involved only in the optimality part, not the uniqueness part.
\end{IEEEproof}

Next, we show that $Ax-b$ is constant for $x$ over $X_\lambda,Y_\sigma,Z_\tau$, and we first prove a simple  lemma.
\begin{lemma}\label{lem01}
Let $f$ be a strictly convex function. If $f(Ax-b)+\|x\|_1$ is constant on a convex set $S$, then both $Ax-b$ and $\|x\|_1$ are  constant on $S$.
\end{lemma}
\begin{IEEEproof}
It suffices to prove the case where $S$ has more one point. Let $x_1$ and $x_2$ be any two  different points in $S$. Consider the line segment $L$ connecting $x_1$ and $x_2$. Since  $X$ is convex, we have $L\subset X$ and that   $f(Ax-b)+\|x\|_1$ is constant on $L$.  On one hand, $\|x\|_1$ is piece-wise linear over $L$; on the other hand, the strict convexity of $f$ makes it impossible for $f(Ax-b)$ to be piece-wise linear over $L$ unless $Ax_1-b=Ax_2-b$. Hence, we have $Ax_1-b=Ax_2-b$ and thus $f(Ax_1-b)=f(Ax_2-b)$, from which it follows  $\|x_1\|_1= \|x_2\|_1$. Since $x_1$ and $x_2$ are arbitrary two points in $S$, the lemma is proved.
\end{IEEEproof}
With Lemma \ref{lem01} we can show
\begin{lemma}\label{lem02}
Under Assumptions \ref{assmp2} and \ref{assmp3}, the following statements for problems \eqref{eq:qpa}--\eqref{eq:qpc} hold
\begin{itemize}
\item[1)]  $X_\lambda, Y_\sigma$ and $Z_\tau$ are convex;
\item[2)]  In problem \eqref{eq:qpa}, $Ax-b$ and $\|x\|_1$ are constant for all $x\in X_\lambda$;
\item[3)]  Part 2) holds for problem \eqref{eq:qpb} and $Y_\sigma$;
\item[4)]  Part 2) holds for problem \eqref{eq:qpc} and $Z_\tau$.
\end{itemize}
\end{lemma}
\begin{IEEEproof}
Assumption \ref{assmp2} makes sure that $X_\lambda,Y_\sigma,Z_\tau$ are all non-empty.

1) As a well-known result, the  set of solutions of a convex program is convex.

2) Since $f_1(Ax-b)+\lambda \|x\|_1$ is constant over $x\in X_\lambda$ and $f_1$ is strictly convex by Assumption \ref{assmp3}, the result follows directly from Lemma \ref{lem01}.

3) If $\vzero\in Y_\sigma$, then the optimal objective is $\|\vzero\|_1= 0$; hence, $Y_\lambda=\{\vzero\}$ and the results hold trivially. Suppose $\vzero\not\in Y_\sigma$. Since the optimal objective $\|x\|_1$ is constant for all $x\in Y_\sigma$ and $f_2$ is strictly convex by Assumption \ref{assmp3}, to prove this part in light of Lemma \ref{lem01}, we shall show $f_2(Ax-b)=\sigma$ for all $x\in Y_\sigma$.

Assume that there is $\hat{x}\in Y_\sigma$ such that $f_2(A\hat{x}-b)<\sigma$. Since $f_2(Ax-b)$ is convex and thus continuous in $x$, there exists a non-empty ball $\cB$ centered at $\hat{x}$ with a sufficiently small radius $\rho>0$ so that $f_2(A\bar{x}-b)<\sigma$ for all $\bar{x}\in\cB$.  Let $\alpha = \min\{\frac{\rho}{2\cdot \|\hat{x}\|_2},\frac{1}{2}\}\in (0,1)$. We have $(1-\alpha)\hat{x}\in\cB$ and  $\|(1-\alpha)\hat{x}\|_1= (1-\alpha)\|\hat{x}\|_1<\|\hat{x}\|_1$, so $(1-\alpha)\hat{x}$ is both feasible and achieving an objective value lower than the optimal one. Contradiction.

4) By Assumption \ref{assmp3}, we have $\|x\|_1=\tau$ for all $x\in Z_\tau$; otherwise, there exists $\bar{x}\in Z_\tau$ such that  $\tau>f_3(A\bar{x}-b)\ge \inf \{\|x\|_1: f_3(Ax-b) = f_3^*\}$, contradicting Assumption \ref{assmp3}. Since the optimal objective $f_3(Ax-b)$ is constant for all $x\in Z_\tau$ and $f_3$ is strictly convex by Assumption \ref{assmp3},  the result follows from Lemma \ref{lem01}.
\end{IEEEproof}

\begin{IEEEproof}[Proof of Theorem \ref{thm:unique} for problems \eqref{eq:qpa}--\eqref{eq:qpc}]
This proof exploits Lemma \ref{lem02}. Since the results of Lemma \ref{lem02} are are identical for problems \eqref{eq:qpa}--\eqref{eq:qpc}, we  present the proof  for problem \eqref{eq:qpa}. The proofs for the other two problems are similar.

From Assumption \ref{assmp3}, $X_\lambda$ is nonempty so we pick $x^*\in X_\lambda$. Let $b^* = Ax^*$, which is independent of the choice of $x^*$ according to Lemma \ref{lem02}. We introduce the linear program
\beq\label{eq:dd}
\min \|x\|_1,\quad\text{s.t.}~Ax= b^*,
\eeq
and let $X^*$ denote its solution set.

Now, we show that $X_\lambda = X^*$. Since $Ax = Ax^*$ and $\|x\|_1 = \|x^*\|_1$ for all $x\in X_\lambda$ and conversely any $x$ obeying $Ax=Ax^*$ and $\|x\|_1=\|x^*\|_1$  belongs to $X_\lambda$, it is suffices to show that $\|{x}\|_1 = \|x^*\|_1$ for any ${x}\in X^*$. Assuming this does \emph{not} hold, then since problem \eqref{eq:dd}  has  $x^*$ as a feasible solution and has a finite objective,  we have a nonempty $X^*$ and there exists $\bar{x}\in X^*$ satisfying $\|\bar{x}\|_1 < \|x^*\|_1$. But,  $f(A\bar{x}-b) = f(b^*-b)=f(Ax^*-b)$ and $\|\bar{x}\|_1 < \|x^*\|_1$ mean that $\bar{x}$ is a strictly better solution to problem \eqref{eq:qpa} than $x^*$, contradicting the assumption $x^*\in X_\lambda$.

Since $X_\lambda = X^*$,  $x^*$ is the unique solution to problem \eqref{eq:qpa} if and only if it is the same to problem \eqref{eq:dd}. Since problem \eqref{eq:dd} is in the same form of problem \eqref{eq:bp}, applying the part of Theorem \ref{thm:unique} for problem \eqref{eq:bp}, which is already proved, we conclude that  $x^*$ is the unique solution to problem \eqref{eq:qpa} if and only if Condition \ref{cond:unique} holds.
\end{IEEEproof}
\begin{IEEEproof}[Proof of Theorem \ref{thm:qp}] The proof above also serves the proof for Theorem \ref{thm:qp} since \eqref{eq:opta}--\eqref{eq:optc} are the optimality conditions of $x^*$ to problems \eqref{eq:qpa}--\eqref{eq:qpc}, respectively, and furthermore,  given the optimality of $x^*$, Condition \ref{cond:unique} is the necessary and sufficient condition for the uniqueness of $x^*$.
\end{IEEEproof}
\begin{remark}
For problems \eqref{eq:qpa}--\eqref{eq:qpc}, the uniqueness of a given solution $x^*\not=\vzero$ is also equivalent to a condition that is slightly simpler than Condition \ref{cond:unique}. To present the condition, consider the first-order optimality conditions (the KKT conditions) \eqref{eq:opta}--\eqref{eq:optc} of $x^*$ to problems \eqref{eq:qpa}--\eqref{eq:qpc}, respectively,
Given $x^*\not=\vzero$, $\eta$ and $\nu$ can be computed. From $p^*\not=\vzero$ it follows that $\eta>0$. Moreover, $\nu=0$ if and only if $A^T \nabla f_3(Ax^*-b)=\vzero$. The condition below for the case $\nu =0$ in problem \eqref{eq:qpc} reduces to Condition \ref{cond:unique}. Define
\begin{eqnarray}\nonumber
&P_1&=  \{i: |\lambda a^T_i \nabla f_1(Ax^*-b)|=1\},\\ \nonumber
&P_2&=  \{i:|\eta a_i^T \nabla f_2(Ax^*-b)|=1\},\\ \nonumber
&P_3&=  \{i:| a_i^T \nabla f_2(Ax^*-b)|=\nu\}. \nonumber
\end{eqnarray}
By the definitions of $\partial\|x^*\|_1$ and $P_i$, we have $\supp(x^*)\subseteq P_i$, $i=1,2,3$.
\begin{condition}\label{cond:sub} Under the definitions  $I:=\supp(x^*)\subseteq P_i$ and $s:=\sign(x_I^*)$, matrix $A_{P_i}\in\mathbb{R}^{m\times |P_i|}$ obeys
\begin{enumerate}
\item submatrix $A_{I}$ has full column rank, and
\item there exists  $y \in \mathbb{R}^m$ such that  $A_{I}^Ty=s$ and $\|A_{P_i\setminus I}^Ty\|_\infty <1$.
\end{enumerate}
Compared to Condition \ref{cond:unique}, Condition \ref{cond:sub} only checks the submatrix $A_{P_i}$ but not the full matrix $A$.
\end{condition}
It is not difficult to show that the linear programs
$$\min \|x\|_1,\quad\text{s.t.}~(A_{P_i})x= b^*,$$
for $i=1,2,3$, have the solution sets that are equal to $X_\lambda,Y_\sigma,Z_\tau$, respectively. From this argument, we have
\begin{theorem}
Under Assumptions \ref{assmp1}--\ref{assmp3} and the additional condition that $f_1,f_2,f_3\in C^1$, given that $x^*\not=\vzero$ is a solution  to problem \eqref{eq:qpa}, \eqref{eq:qpb}, or \eqref{eq:qpc},  $x^*$ is  the unique solution if and only if Condition \ref{cond:sub} holds for $i=1$, $2$, or $3$, respectively.
\end{theorem}
\end{remark}
\begin{IEEEproof}[Proof of Corollary \ref{corol}]
From Theorem \ref{thm:qp}, if $x^*$ is the unique minimizer of problem \eqref{eq:qpa}  with $f_1(\cdot)=\frac{1}{2}\|\cdot\|_2^2$, then Condition 1 holds, so there must exist a vector $y$ such that  $A_{I}^Ty=s$ and $\|A_{I^c}^Ty\|_\infty <1$. Combining with (\ref{cond2a}),  we have
$$\lambda A_{I}^Ty=\lambda  s=A_{I}^T(b-A_{I}x^*).$$
Since $A_{I}$  is a full rank square matrix, we get $y=\frac{1}{\lambda} (b-A_{I}x^*)$. Substituting this formula to $\|A_{I^c}^Ty\|_\infty <1$, we obtain condition (\ref{cond0}).

\end{IEEEproof}

\begin{IEEEproof}[Proof of Corollary \ref{coro2}]
The sufficiency part follows from Theorem \ref{Ti12}. We shall show the necessity part, namely,  if $x^*$ is the unique solution, then $A_J$ has full column rank. Following the assumption $|J|=|I|+1$, we let $\{i_0\}=  J\backslash I$. Since $x^*$ is the unique solution, from Theorem \ref{thm:qp}, we know that $A_I$ has full column rank. Hence, if $A_J$ does \emph{not} have full column rank, then we can have $a_{i_0}=A_I \beta$ for some $\beta\in \RR^{|I|}$. From Theorem \ref{thm:qp}, if $x^*$ is the unique minimizer,  then Condition \ref{cond:unique} holds, and in particular, there must exist a vector $y$ such that  $A_{I}^Ty=s$ and $\|A_{I^c}^Ty\|_\infty <1$. Now, on one hand, as $i_0\in I^c$, we get $1> |a_{i_0}^Ty|=  |\beta^TA_I^Ty| =|\beta^Ts|$; on the other hand,  as $i_0\in J$, we also have $|a_{i_0}^T(b-Ax^*)|=\lambda$, which implies $1=\frac{1}{\lambda}|a_{i_0}^T(b-Ax^*)|=\frac{1}{\lambda} |\beta^TA_I^T(b-Ax^*)|=|\beta^T s|$, where the last equality follows from \eqref{eq:opta} (which includes \eqref{cond2a}). Contradiction.
\end{IEEEproof}

\section{Recognizing and verifying unique solutions}
Applying Theorem \ref{thm:unique}, we can recognize the uniqueness of a given  solution $x^*$ to problem \eqref{eq:bp} given a dual solution $y^*$ (a solution to problem \eqref{eq:bpdual}). In particular, let $J:=\{i:|a^T_i y^*|=1\}$, and if $A_J$ has full column rank and $\supp(x^*)=J$, then according to Theorem \ref{thm:unique}, $x^*$ is the unique solution to \eqref{eq:bp}. The converse  is not true since there generally exist multiple dual solutions which have different $J$. However, several linear programming interior point methods (see \cite{GulerYe} for example) return the dual solution $y^*$ with the smallest $J$, so if either $A_J$ is column-rank deficient or $\supp(x^*)\not=J$, then $x^*$ is surely non-unique.
\begin{corollary} Under Assumption \ref{assmp1}, given a pair of primal-dual solutions $(x^*,y^*)$ to problem \eqref{eq:bp}, let $J:=\{i:|a^T_i y^*|=1\}$. Then, $x^*$ is the unique solution to \eqref{eq:bp} if $A_J$ has full column rank and  $\supp(x^*)=J$. In addition, if $y^*$ is obtained by a linear programming interior-point algorithm, the converse also holds.
\end{corollary}
Similar results will also hold for problems \eqref{eq:qpa}--\eqref{eq:qpc} if a dual solution $y^*$ to \eqref{eq:dd} is available.

One can also directly verify Condition \ref{cond:unique}. Given a matrix $A\in\RR^{m\times n}$, a set of its columns indexed by $I$, and a sign pattern $s=\{-1,1\}^{|I|}$, we mention two approaches to verify Condition \ref{cond:unique}. Checking whether $A_I$ has full column rank is straightforward. To check part 2 of Condition \ref{cond:unique}, the first approach is to follow the proof of Theorem \ref{thm:unique}. Note that Condition \ref{cond:unique} depends only on $A$, $I$, and $s$, independent of $x^*$. Therefore, construct an \emph{arbitrary} $x^*$ such that $\supp(x^*)=I$ and $\sign(x_I^*)=s$ and let $b=Ax^*$. Solve problem \eqref{eq:bpdual} and let $y^*$ be its solution. If $y^*$ satisfies part 2 of Condition \ref{cond:unique}, we are done; otherwise, define $L$, $U$, and $t$ by $x^*$ as in the proof, pick a small $\bar{\alpha}>0$, and solve  program \eqref{eq:idual} parametrically in $\alpha\in [0,\bar{\alpha}]$. The solution is piece-wise linear in $\alpha$ (it is possible that the solution does not exist over certain intervals of $\alpha$). Then check if  there is a perturbation to $y^*$ so that $y^*$ satisfies part 2 of Condition \ref{cond:unique}.

In the second approach to check part 2 of Condition \ref{cond:unique}, one can solve
 the  convex program
\begin{equation}\label{eq:check}
\begin{split}
&\min_{y\in\RR^m}-\sum_{i\in I^c} \log(1 - a_i^Ty) +\log(1 + a_i^Ty),\\
&\quad\text{s.t.}~A_I^T y = s.
\end{split}
\end{equation}
Since $a_i^Ty\to 1$ or $a_i^Ty\to -1$ will infinitely increase the objective, \eqref{eq:check} will return a solution satisfying Condition \ref{cond:unique}, part 2, as long as a solution exists. In fact, any feasible solution to  \eqref{eq:check} with a finite objective satisfies Condition \ref{cond:unique}, part 2. To find a feasible solution, one can apply the augmented Lagrangian method, which does not require $A_I^T y = s$ to hold at the initial point (which must still satisfy $|a_i^Ty|<1$ for all $i\in I^c$), or one can consider applying the alternating direction method of multipliers (ADMM) to the equivalent problem
\begin{equation}\label{eq:check1}
\begin{split}
&\min_{y,z}-\sum_{i\in I^c} \log(1 - z_i) +\log(1 + z_i),\\
& \quad\text{s.t.}~A_I^T y = s,~z -A^T_{I^c}y=\vzero.
\end{split}
\end{equation}
One can start ADMM from the origin, and the two subproblems of ADMM have closed-form solutions; in particular, the $z$-subproblem is separable in $z_i$'s and reduces to finding the zeros of 3-order polynomials in $z_i$, $i\in I^c$. If \eqref{eq:check1} has a solution, ADMM will find one; otherwise, it will diverge.

{It is worth mentioning that one can use alternating projection in \cite{DL} to generate test instances that fulfill Condition \ref{cond:unique}.}

\section{Conclusions}
This paper shows that Condition \ref{cond:unique}, which is previously known to be sufficient for the solution uniqueness of the basis pursuit model, is also necessary. Moreover, the condition applies to various $\ell_1$ minimization models. The result essentially follows from the fact that a pair of feasible primal-dual programs have   strict complementary solutions. The result also sheds light on numerically recognizing unique solutions and verifying solution uniqueness.

\section*{Acknowledgements}
The authors thanks Prof. Dirk Lorenz for bringing references \cite{GHS} and \cite{DL} to their attention. H. Zhang thanks Rice University, CAAM Department, for hosting him.

%\begin{IEEEbiography}{Wotao Yin}
%\end{IEEEbiography}

%\begin{IEEEbiographynophoto}{WotaoYin}


\begin{thebibliography}{1}
\bibitem{BG} M. J. Best, and R. R. Grauer, ``Sensitivity analysis for mean-variance portfolio problems,"  Management Science, vol. 37, no. 8, pp. 980-990, 1991.

\bibitem{BDE}A. M. Bruckstein,  D. L. Donoho, and M. Elad, ``From Sparse Solutions of Systems of Equations to Sparse Modeling of Signals and Images,"   SIAM Review, vol. 51, no. 1, pp. 34-81, 2009.

\bibitem{CT}E. J. Cand\`{e}s and T. Tao,  ``Decoding by linear programming,"   IEEE Trans. Inform. Theory,  vol. 51, pp. 4203-4215, 2005.

\bibitem{CRT}E. J. Cand\`{e}s, J. Romberg and T. Tao, ``Robust uncertainty principles: exact signal reconstruction from highly incomplete frequency information,"   IEEE Trans. Inform. Theory, vol. 52, no. 2, pp. 489-509, 2006.

%\bibitem{CR09}E. J. Cand\`{e}s and B. Recht. Exact matrix completion via convex optimization, Foundations of Computational Mathematics, 9:717-772, 2009.

%\bibitem{CLMW}E. J. Cand\`{e}s, X. Li, Y. Ma and J. Wright. Robust principal component analysis?, Journal of ACM, 58:1-37, 2011.

\bibitem{CP11}E. J. Cand\`{e}s and Y. Plan, ``A probabilistic and RIPless theory of compressed sensing,"   IEEE Trans. Inform. Theory, vol. 57, no. 11, pp. 7235-7254,  2011.

\bibitem{CR11}E. J. Cand\`{e}s and B. Recht, ``Simple bounds for low-complexity model reconstruction,"   To appear in Math. Program.,  2012.

\bibitem{Da}G. B. Dantzig, Linear Programming and Extensions, Princeton, 1963.

%\bibitem{CRPW}V. Chandrasekaran, B. Recht, P. A. Parrilo, and A. Willsky. The convex geometry of linear inverse problems. Submitted for publication, Submitted to Foundations of Computational Mathematics, Preprint available at arxiv.org/1012.0621, 2010.

\bibitem{CDS}S. S. Chen, D. L. Donoho, and M. A. Saunders, ``Atomic decomposition by basis pursuit,"   SIAM J. Sci. Comput., vol. 20,  pp. 33-61, 1999.

\bibitem{Co}A. Cohen, W. Dahmen, and R. DeVore, ``Compressed sensing and best k-term approximation,"   J. Amer. Math. Soc., vol. 20,  pp. 211-231, 2009.

\bibitem{DE}D. L. Donoho and M. Elad, ``Optimally sparse representation in general(non-orthogonal) dictionaries via $\ell_1$ minimization,"   Proc. Natl. Acad. Sci.,  vol. 100, no. 5,  pp. 2197-2202, 2003.

\bibitem{DH}D. L. Donoho and X. Huo, ``Uncertainty principles and ideal atomic decomposition,"   IEEE Trans. Inform. Theory, vol. 47, no. 7, pp. 2845-2862, 2001.

\bibitem{Do06}D. L. Donoho, ``Compressed sensing,"   IEEE Trans. Inform. Theory, vol. 52, no. 4, pp.1289-1306, 2006.

\bibitem{Do}C. Dossal, ``A necessary and sufficient condition for exact sparse recovery by $\ell_1$ minimization,"   C. R. Acad. Sci. Paris, Ser. I, vol. 350, pp. 117-120, 2012.

\bibitem{EJHT}B. Efron, I. Johnstone, T. Hastie,  and R. Tibshirani, ``Least angle regression,"   Ann. of Statist., vol. 32, pp. 407-499, 2004.

\bibitem{EB}M. Elad and A. M. Bruckstein, ``A generalized uncertainty principle and sparse representation in pairs of bases,"   IEEE Trans. Inform. Theory, vol. 48, no. 9, pp. 2558-2567,  2002.

\bibitem{Fu}J. J. Fuchs, ``On sparse representations in arbitrary redundant bases,"  IEEE Trans. Inform. Theory,  vol. 50, no. 6, pp.1341-1344, 2004.

\bibitem{Fu05}J. J. Fuchs, ``Recovery of Exact Sparse Representations in the Presence of Bounded Noise,"   IEEE Trans. Inform. Theory,  vol. 51, no. 10, pp. 3601-3608, 2005.

\bibitem{GHS}M. Grasmair, M. Haltmeier, and O. Scherzer, ``Necessary and Sufficient Conditions for Linear Convergence of $\ell_1$-Regularization," Comm. Pure
Appl. Math., vol. 64, no. 2, pp.161-182, 2011.

\bibitem{GulerYe} O. G\"uler and Y. Ye,  ``Convergence behavior of interior-point algorithms," Math. Program.,  vol. 60, pp. 215-228, 1993.

\bibitem{DL}D. Lorenz, ``Constructing test instances for Basis Pursuit Denoising,"
arXiv.org/abs/1103.2897, 2011.

\bibitem{Ti}R. Tibshirani, ``Regression shrinkage and selection via the lasso,"   J. R. Statist. Soc. B,  vol. 58, no. 1, pp. 267-288, 1996.

\bibitem{Ti12}R. J. Tibshirani, ``The Lasso problem and uniqueness,"  arXiv:1206.0313v1,  2012.

\bibitem{Tr}J. A. Tropp,  ``Recovery of short, complex linear combinations via $\ell_1$ minimization,"   IEEE Trans. Inform. Theory, vol. 51, no. 4, pp. 1568-1570,  2005.

\bibitem{Zh}Y. Zhang, Theory of compressive sensing via $\ell_1$-minimization: a non-RIP analysis and extensions,  Rice University, Houston, TX, Tech. Rep., 2008.
\end{thebibliography}
\end{document}